\documentclass[twocolumn,showpacs,preprintnumbers,amsmath,amssymb]{revtex4}
\usepackage{tabularx,graphicx}\begin{document}
\newcommand{\beq}{\begin{equation}}
\newcommand{\eeq}{\end{equation}}
\newcommand{\beqn}{\begin{eqnarray}}
\newcommand{\eeqn}{\end{eqnarray}}
\newcommand{\bmath}{\begin{subequations}}
\newcommand{\emath}{\end{subequations}}
\title{Explanation of the Tao effect}
\author{J. E. Hirsch }
\address{Department of Physics, University of California, San Diego\\
La Jolla, CA 92093-0319}
 
\date{January 2, 2005} 
\begin{abstract} 
In a series of experiments Tao and coworkers\cite{tao1,tao2,tao3} found that superconducting microparticles in the
presence of a strong electrostatic field aggregate into balls of macroscopic dimensions.   No explanation of this 
phenomenon exists within the conventional theory of superconductivity. We show that this effect
can be understood within an alternative electrodynamic description of superconductors recently proposed that
follows from an unconventional theory of superconductivity.  
Experiments to test the theory are discussed.

\end{abstract}
\maketitle 

The phenomenon was discovered by Tao and coworkers in 1999 in samples of high $T_c$ superconducting
materials, and initially attributed to special properties of high $T_c$ cuprates, in particular
their layered structure\cite{tao1}. However, subsequent experiments reported in 2002 for 
conventional superconducting materials ($Pb$, $V$, $V_3Ga$, $Nb-N$ and $Nb_3Sn$) all showed
the same behavior\cite{tao2}, and the same behavior was observed in $MgB_2$\cite{tao3}. Briefly the 
remarkable observation is: when the applied electric field exceeds a critical value, typically of order 0.5-1 kV/mm, millions of 
superconducting microparticles each of size $\sim 5\mu$ spontaneously aggregate into spherical 
balls of size of order $mm$. As the electric field increases further the size of the balls decreases, and above a second critical field,
of order 1.5-2 kV/mm the balls disintegrate and the microparticles fly to the electrodes and cling to them. 
The conventional theory of superconductivity predicts that superconductors
respond to applied electrostatic fields in the same way as normal metals do\cite{london,bardeen}.
Because normal metallic microparticles do not aggregate into spherical balls upon application
of electrostatic fields nor do they cling to the electrodes, Tao's observation represents a fundamental 
puzzle within the conventional understanding of superconductivity. For high $T_c$ materials, Tao's findings
have been independently confirmed\cite{ryle}

We have proposed an unconventional theory of superconductivity to describe 
both high $T_c$\cite{hole1} and conventional
superconductors\cite{hole2} that is based on the fundamental charge asymmetry of 
matter\cite{hole3}.
This theory leads to a new description of the electrodynamic properties of superconductors\cite{electro}.
Here we show that this new formulation provides an explanation for the Tao effect.

In this theoretical framework, negative charge is expelled from the interior of the material
towards the surface when the transition to superconductivity occurs.
The resulting charge density obeys the differential equation
\beq
\rho(\vec{r})=\rho_0+\lambda_L^2\nabla^2\rho(\vec{r})
\eeq
in the interior of the superconductor, with $\lambda_L$ the London penetration depth
and $\rho_0$ a $positive$ constant that is function of the parameters of the 
material and of the dimensions of the sample. Eq. (1) predicts that the charge density
is $\rho_0$ deep in the interior of the superconductor. By charge neutrality
the average charge density is negative near the surface ($=\rho_-$) and is approximately
related to $\rho_0$ by $\rho_-S\lambda_L=-\rho_0 V$
with $S$ and $V$ the surface area and volume of the body. Energetic arguments show that
$\rho_-$ is independent of the volume of the sample for samples of dimensions much larger
than the London penetration depth. The electrostatic potential $\phi$ obeys the equation
\beq
\phi (\vec{r})=-4\pi \lambda_L^2 \rho (\vec{r})+\phi _0(\vec{r})
\eeq
in the interior of the superconductor, with $\phi _0$ the electrostatic potential due to a uniform
positive charge density $\rho_0$. Justification of these equations is given in Ref.\cite{electro}.  
 In the exterior, $\phi (\vec{r})$ 
obeys Laplace's equation $\nabla^2 \phi =0$, and furthermore we assume that
$\phi $ and its normal derivative are continuous at the surface of the superconductor,
hence that no surface charge exists.

Under these conditions a unique solution for the charge distribution and electrostatic potential
exists for given sample geometry. For samples of high symmetry (eg spherical,
infinite cylinder or infinite plane) no electric field exists in the exterior of the 
superconductor. However for samples of general shape these equations predict that
'spontaneous' electric fields exist outside the sample near the surface.

Consider samples of ellipsoidal shape. Figure 1 shows the electric field lines obtained for prolate and
oblate ellipsoids of eccentricity $|e|=0.745$, corresponding to ratio of axis $b/a=1.5$ and $a/b=1.5$ 
respectively,  and London penetration depth
$\lambda_L=0.2$ in units where $a^2b=1.5$. For other values of 
the penetration depth the results are very similar. In both cases electric field lines go out in the region of low surface curvature
and go in in regions of high surface curvature. We find the same behavior in samples
of other shapes. The magnitude of these quadrupolar electric fields increases as the
eccentricity increases.
Examples of the charge distribution inside the superconductor that gives
rise to these electric fields are given in ref.\cite{ellip}. The reason that macroscopic charge inhomogeneity and
differences in electric potential in the  interior of superconductors can exist is because
superfluid electrons will compensate for the difference in $potential$ $energy$ in different regions with
corresponding changes in their $kinetic$ $energy$\cite{electro} . 

 \begin{figure}
\resizebox{8.0cm}{!}{\includegraphics[width=7cm]{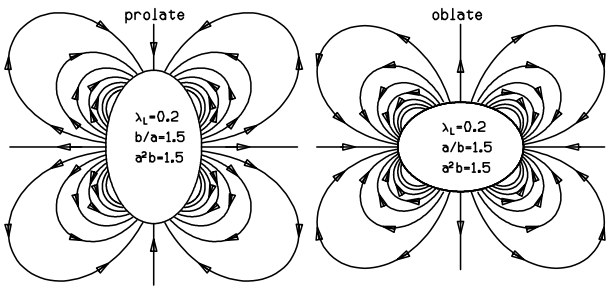}}
\caption{Electric field line configurations for a prolate and an oblate ellipsoid, with equatorial radius a and polar radius b.
}
\label{Fig. 1}
\end{figure}

Figure 2 shows the behavior of the electrostatic potential on the surface of the ellipsoids of Fig. 1.
In the prolate (oblate) case the maximum potential occurs at the equator (poles). Fig. 2 also shows
the potential for a quadrupole moment at the origin of the magnitude obtained from integration over
the interior charge distribution 
\beq
Q=\int_V d^3r \rho(\vec{r}) (3z^2-r^2)
\eeq
At distances further away the actual potential rapidly approaches the one generated by a pure quadrupole of
value given by Eq. (4), as seen in Fig. 2.

We calculate the electrostatic energy associated with these electric fields and charge distribution
\beq
U_E=\int d^3r \frac{|\vec{E}|^2}{8\pi}=\frac{1}{2}\int_V d^3r \rho(\vec{r})\phi(\vec{r})
\eeq 
with $\vec{E}=-\vec{\nabla}\phi$. The first integral is over all space, the second over the volume of the sample.
The contribution to the electrostatic energy associated with fields $outside$ the sample is given by
the surface integral
\beq
U_{out}=-\int_S dS  \frac{\partial \phi}{\partial n} \phi
\eeq
and the energy from the electric field
 inside of the sample is $U_{in}=U_E-U_{out}$. We assume that for samples of
different shapes and fixed volume, $\rho_-$ is constant, and calculate these energies for ellipsoids of
fixed volume and varying eccentricity. 
 Results are shown in Fig. 3.  The minimum energy occurs for a spherical shape, 
where only an electric field in the
interior exists. As the ellipsoid deviates from spherical shape the energy due to the outside electric fields
increases rapidly, while the one from the interior electric field decreases somewhat, giving rise to an
increase in the total electrostatic energy which goes linearly with the increase in surface area of the
sample and coefficient close but somewhat larger than unity.
 \begin{figure}
\resizebox{8.5cm}{!}{\includegraphics[width=7cm]{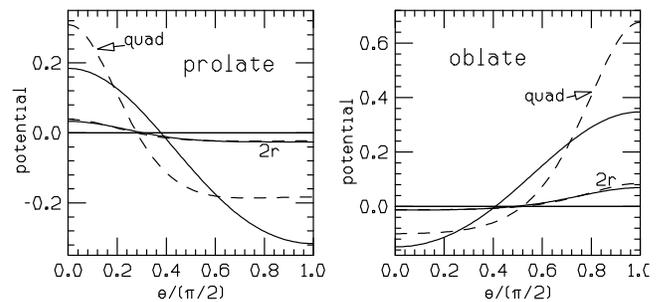}}
\caption{
Electrostatic potential on the surface of the ellipsoid versus angle with the equatorial direction $\theta$. The dashed lines (labeled 'quad')
      show the potential generated by a pure quadrupole  Q=-1.24 (Q=0.902) for the prolate (oblate) case. The almost-coincident 
     solid and dashed curves labeled '2r' give the
     potential and that of the pure quadrupole at twice the distance from the origin.
}
\label{Fig. 2}
\end{figure}

 \begin{figure}
\resizebox{6.5cm}{!}{\includegraphics[width=7cm]{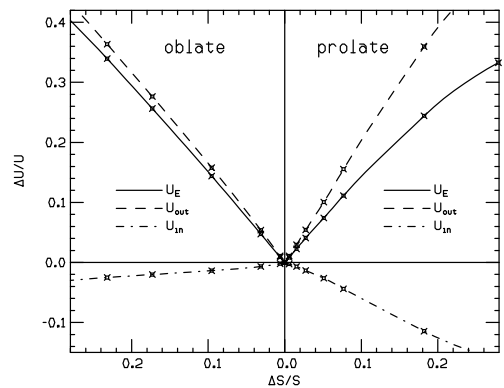}}
\caption{
Fractional electrostatic energy change versus fractional change in surface area relative to the sphere for prolate and oblate
       ellipsoids of fixed volume with $a^2b=1.5$ and $\lambda_L=0.2$, for constant $\rho_-$. The full, dashed and dot-dashed line give
       the total energy $U_E$, energy outside $U_{out}$ and energy inside the ellipsoid $U_{in}$. }
\label{Fig. 3}
\end{figure}

We infer from these results that there is an energetic advantage for superconducting
bodies to  adopt a spherical shape, in order to eliminate the electric fields outside the sample that
raise the electrostatic energy. This is similar to the effect of the surface tension in
liquids, which causes liquid drops to be spherical.   However because
superconducting bodies unlike liquid drops have a rigid structure they cannot
spontaneously deform into spherical shapes.

Consider now the effect of an applied electrostatic field. The phenomenon
of electric field induced coalescence of conducting or dielectric fluid droplets is well known\cite{coal,coal2}. For such liquid drops in
suspension, application of a uniform electrostatic field causes an induced dipole moment in droplets and
an attraction between droplets along the field direction. Upon contact, the surface tension deforms the
elongated droplet into a spherical droplet of larger size than the original ones.

We can envisage a similar scenario for superconducting microparticles. Figure 4a shows four 
spherical microparticles alligned due to the application of an electrostatic field. When they come into
contact, charge  will redistribute to conform to the new shape,
 and the electrostatic energy will increase due to the generation of the 
external electric fields. We can approximate the
resulting electric field by that corresponding to a prolate ellipsoid of aspect ratio $b/a=4$, the
resulting electric field lines are shown in Fig. 4a. If two such groups coalesce to form a 
$2\times2\times 2$ arrangement, as shown schematically in Fig. 4b, the shape is almost spherical and the
electric field outside is nearly zero. The electrostatic energy of the elongated shape is over $30\%$ larger than that of
the spherical shape according to Fig. 3,  so the latter one will be favored.

 \begin{figure}
\resizebox{5.5cm}{!}{\includegraphics[width=7cm]{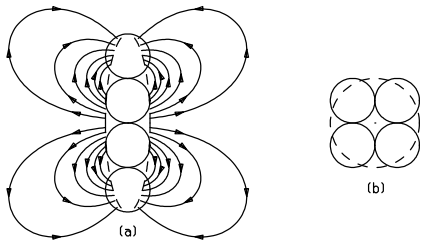}}
\caption{
When 4 spheres of equal radius allign, the result is approximately a prolate ellipsoid of aspect
       ratio $b/a=4$, with a $28\%$ larger surface area than the sphere of equal volume.}
\label{Fig. 4}
\end{figure}

This physics clearly favors clustering of microparticles into spherical shapes as seen in the Tao experiments,
aided by an applied electric field that will induce dipole moments on the microparticles causing them first to allign
as in Fig. 4a.  However we believe there is more to the Tao phenomenon than this. Note that in the electrical coalescence of fluid droplets
the size distribution changes so that the average drop size increases somewhat\cite{coal2}, however there is not an aggregation
of a large number of droplets into a single large drop. We believe the much stronger Tao phenomenon requires the
existence of an actual {\it force between particles that are not in contact} that can act not only in direction
$parallel$ to the applied electric field (as occurs due to the induced electric dipoles) but also in direction
$perpendicular$ to the field, so as to favor a compact arrangement. We now show that such a force indeed
exists in our case.

In fact, the first question to ask is: why doesn't aggregation of microparticles occur even in the absence of 
an applied electric field? The predicted electrostatic fields around nonspherical particles shown in Fig. 1 should give
rise to an attractive force between close-by microparticles in the proper relative orientation, namely high curvature region of one
particle close to low curvature region of another. Only if the microparticles are perfectly spherical will
no electric field exist in the exterior, so why then don't microparticles of random shapes
spontaneously aggregate  to form spherical arrangements?

We believe the answer to this question is that in fact in the absence of an
applied electric field a microparticle cannot sustain differences in electrostatic potential between two surface
points larger than the work function of the material.
If such large potential difference 
on the surface exists
 it will become energetically
advantageous for electrons to 'pop out' of the superconducting condensate in the region of low electric potential and
migrate to the region of high potential, and the resulting electronic layer $outside$ the superconductor will
screen the outgoing electric field lines.  This then implies that   electric field lines that start and end at points on the surface with electric potential difference larger
than the work function, typically a few electron volts, will be screened. We estimated the magnitude of the electric fields
near the surface of superconductors
 at about $10^5 V/cm$\cite{electro}, which for a microparticle of dimension $5\mu$ and eccentricity as
 in Fig. 1 would give a maximum potential
difference between points 
on the surface of about $25V$, much larger than the work function.
 We conclude that such particles will be 'coated' by a layer of charge on the surface
so that the electric field in the exterior will be screened out. 

We propose then that the role of the applied electric field in Tao's experiment
 is to remove the electronic  'coating' that hides these electric fields so that
they become observable. The coating electrons   have already payed the work function price, i.e. they
are outside the surface 'double layer' that gives rise to the work function potential. Hence they can be removed by an
applied electric field strong enough to overcome the force due to the electric field of the superconductor. 
As these coating electrons are removed, the electric field lines shown in Fig. 1 start to become visible and
give rise to forces between microparticles.

We assume then that the applied electric field has removed the electronic coating, and calculate the total electric field
around a microparticle, which is a superposition of the external plus internal field plus the field generated by the
electric dipole that is induced as for a normal metallic particle.   The differential equation to be solved is still
Eq. (2), except that now the boundary condition for the potential changes.
The electrostatic potential is given by
\beq
\phi(\vec{r})=-E\vec{z}+\int_V d^3r \frac{\rho(\vec{r'})}{|\vec{r}-\vec{r'}|}
\eeq
with E the applied electric field along the $z$ direction. The numerical procedure is analogous to that
described in Ref. \cite{ellip}.

 \begin{figure}
\resizebox{6.9cm}{!}{\includegraphics[width=7cm]{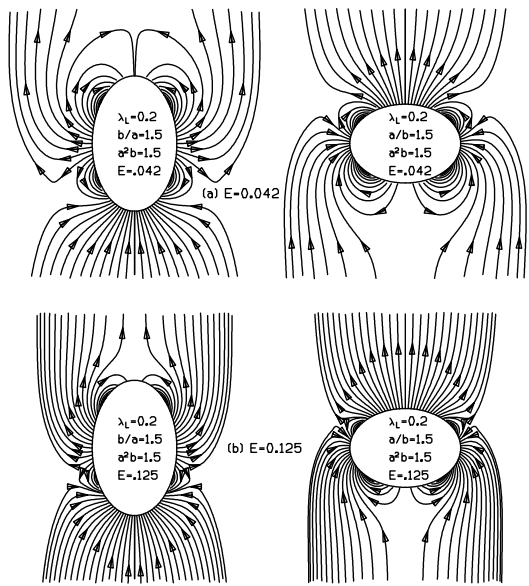}}
\caption{
Electric field lines for  prolate and oblate ellipsoids in the presence of an external uniform electric field E.
}
\label{Fig. 5}
\end{figure}
       
Figure 5 shows electric field lines for a prolate and an oblate ellipsoid in the presence of an applied electric field
in the vertical direction for two values of the applied field strength. In the units used, the maximum spontaneous 
electric field that
we obtain at the surface is about unity; hence the applied fields in Fig. 5 correspond to  420V/mm and 1250 V/mm.
As the electric field increases 
the field lines that start and end on the sample
get eliminated. Note that some electric field lines come out and go in the ellipsoids in directions $perpendicular$ to the
applied field, due to the effect of the spontaneous fields. 
 As a consequence a prolate and an oblate ellipsoid at short
distance from each other arranged in the orientation shown in Fig. 5 will experience a strong attractive
force in direction perpendicular to the applied electric field . An attractive force in direction parallel to the 
applied field also exists as in the
normal case. These forces in combination with the fact that spherical arrangement leads to minimization of the
electrostatic energy as shown above should act to agglomerate particles of random shapes and orientations
into spherical balls as observed by Tao et al. As the field becomes stronger, the field lines that give rise to the
perpendicular forces bend towards the direction of the electrodes and the  forces perpendicular to the direction 
of the applied field will be suppressed, which will weaken the
tendency to spherical aggregation as observed experimentally.

The  magnitude of the perpendicular forces generated   
can be  estimated from the magnitude of the
quadrupole moments, which is of order $1$ in our units for the ellipsoids in Fig. 1.  
$F\sim Q_1 Q_2/d^6 \sim E^2 d^2$, with $d$ the distance between the centers and $E$ the electric fields at the
surface, yields a force between microparticles of order
 $\sim 10^{-2}dyn$. Tao estimated an average acceleration for the balls colliding with
the electrodes of at least $10g$, which for a $5\mu$ microparticle of density $\rho\sim 10g/cm^3$
would be a force of $0.5\times10^{-4}dyn$, much smaller than the force that we estimate to hold the microparticles
together, so indeed the balls should be able to survive such impacts. 

Furthermore, once the applied electric field has removed the 'coating layer' on the surface of the microparticles,
the spontaneous electric field of the microparticles will exert a strong force on electric charges in 
the electrodes, causing microparticles to 'cling' to the electrodes as observed by Tao et al.
This 'clinging' should increase as the applied electric field and hence the charge on the 
electrodes increases and exist also beyond the field strength where the 
aggregation into balls no longer occurs, as observed experimentally.

Experimental test of our scenario should be possible. For particles of sufficiently small size
($<\sim 0.5 \mu$) differences in the electrostatic potential on the surface should be smaller than
the work function, hence screening of the exterior electric fields  should not occur and spontaneous aggregation should
occur in the $absence$ of an applied electric field. For larger particles, it should be possible to measure the
resulting field configuration in the neighborhood of the surface upon application of a strong electric field
that 'uncoats' the particles, and consistency with the predictions of our theory can be checked
by solution of the differential equation (2). In an inhomogeneous applied strong
electric field, the force and torque acting on a particle will undergo distinct changes between the normal and  superconducting state   that can
be calculated for given shape and orientation of the particle.

 Normal metal microparticles under applied electric fields allign in elongated arrangements due to 
induced electric dipoles but do not form spherical balls. The conventional London-BCS theory of superconductivity predicts that the static dielectric response function of superconductors is the same as that of normal metals\cite{bardeen,koyama}, however Tao's experiment shows that the response is in fact qualitatively different. Our alternative theory was shown here to be consistent with Tao's observations. We 
predict a definite relation between the shape of a superconducting particle and
the magnitude $and$ $sign$ of resulting electric fields near its surface, a manifestation of the fundamental charge
asymmetry of matter. 
 
    
    \end{document}